\documentclass[fleqn,usenatbib]{mnras}

\usepackage{newtxtext,newtxmath}

\usepackage[T1]{fontenc}

\usepackage{graphicx}	
\usepackage{amsmath}	
\usepackage{siunitx}
\usepackage[normalem]{ulem}

\title[Bowshocks: AFGL 5142-MM1]{The Limits of Water Maser Kinematics: Insights from High-Mass Protostar AFGL 5142-MM1}

\author[Rosli Z. et al.]{Zulfazli Rosli,$^{1,2}$\thanks{E-mail:zulfazli.rosli@gmail.com}
Ross A. Burns,$^{3}$
Affan Adly Nazri,$^{1}$
Koichiro Sugiyama,$^{4}$
Tomoya Hirota,$^{5,6}$\newauthor
Kee-Tae Kim,$^{7}$
Yoshinori Yonekura,$^{8}$
Liu Tie,$^{9}$
Gabor Orosz,$^{10}$
James Okwe Chibueze,$^{11,12,13}$\newauthor
Andrey M. Sobolev,$^{14}$
Ji Hyun Kang,$^{7}$
Chang Won Lee,$^{7,15}$
Jihye Hwang,$^{7}$
Hafieduddin Mohammad,$^{16}$\newauthor
Norsiah Hashim,$^{17}$ and
Zamri Zainal Abidin$^{1}$\\
$^{1}$Department of Physics, Faculty of Science, Universiti Malaya, 50603, Kuala Lumpur, Malaysia\\
$^{2}$CFLMS, International University of Malaya Wales, Kuala Lumpur, Malaysia\\
$^{3}$RIKEN Cluster for Pioneering Research, 2-1 Hirosawa, Wako-shi, Saitama, 351-0198, Japan\\
$^{4}$National Astronomical Research Institute of Thailand (Public Organization), 260 Moo 4, T. Donkaew, A. Maerim, Chiangmai, 50180 Thailand\\
$^{5}$Mizusawa VLBI Observatory, National Astronomical Observatory of
Japan, Hoshigaoka 2-12, Mizusawa, Oshu, Iwate 023-0861, Japan\\
$^{6}$The Graduate University for Advanced Studies, SOKENDAI, 2-21-1
Osawa, Mitaka, Tokyo 181-8588, Japan\\
$^{7}$Korea Astronomy and Space Science Institute, 776 Daedeokdae-ro, Yuseong-gu, Daejeon 34055, Republic of Korea\\
$^{8}$Center for Astronomy, Ibaraki University, 2-1-1 Bunkyo, Mito, Ibaraki 310-8512, Japan\\
$^{9}$Shanghai Astronomical Observatory, Chinese Academy of Sciences, 80 Nandan Road, Shanghai 200030, People’s Republic of China\\
$^{10}$Joint Institute for VLBI ERIC, Oude Hoogeveensedijk 4, 7991 PD Dwingeloo, The Netherlands\\
$^{11}$Department of Mathematical Sciences, University of South Africa, Cnr Christian de Wet Rd and Pioneer Avenue, Florida Park, 1709, Roodepoort, South Africa\\
$^{12}$Centre for Space Research, Physics Department, North-West University, Potchefstroom 2520, South Africa\\
$^{13}$Department of Physics and Astronomy, Faculty of Physical Sciences, University of Nigeria, Carver Building, 1 University Road, Nsukka 410001, Nigeria\\
$^{14}$Ural Federal University, 19 Mira Street, 620002 Ekaterinburg, Russia\\
$^{15}$University of Science and Technology, Korea (UST), 217 Gajeong-ro, Yuseong-gu, Daejeon 34113, Republic of Korea\\
$^{16}$Department of Astronomy, University of Tokyo, 2 Chome-21 Osawa, Mitaka, Tokyo 181-0015, Japan\\
$^{17}$Mathematics Division, Centre for Foundation Studies in Science, Universiti Malaya, 50603, Kuala Lumpur, Malaysia}
\date{Accepted XXX. Received YYY; in original form ZZZ}

\pubyear{2023} 

\begin{document}
\label{firstpage}
\pagerange{\pageref{firstpage}--\pageref{lastpage}}
\maketitle

\begin{abstract}
Multi-epoch VLBI observations measure 3D water maser motions in protostellar outflows, enabling analysis of inclination and velocity. However, these analyses assume that water masers and shock surfaces within outflows are co-propagating. We compared VLBI data on maser-traced bowshocks in high-mass protostar AFGL 5142-MM1, from seven epochs of archival data from the VLBI Exploration of Radio Astrometry (VERA), obtained from April 2014 to May 2015, and our newly-conducted data from the KVN and VERA Array (KaVA), obtained in March 2016. We find an inconsistency between the expected displacement of the bowshocks and the motions of individual masers. The separation between two opposing bowshocks in AFGL 5142-MM1 was determined to be $\qty{337.17\pm0.07}{mas}$ in the KaVA data, which is less than an expected value of $\qty{342.1\pm0.7}{mas}$ based on extrapolation of the proper motions of individual maser features measured by VERA. Our measurements imply that the bowshock propagates at a velocity of $\qty{24\pm3}{km.s^{-1}}$, while the individual masing gas clumps move at an average velocity of $\qty{55\pm5}{km.s^{-1}}$, i.e. the water masers are moving in the outflow direction at double the speed at which the bowshocks are propagating. Our results emphasise that investigations of individual maser features are best approached using short-term high-cadence VLBI monitoring, while long-term monitoring on timescales comparable to the lifetimes of maser features, are better suited to tracing the overall evolution of shock surfaces. Observers should be aware that masers and shock surfaces can move relative to each other, and that this can affect the interpretation of protostellar outflows.
\end{abstract}

\begin{keywords}
maser - stars: individual: AFGL 5142-MM1 - stars: massive - ISM: jet and outflows.
\end{keywords}



\section{Introduction}

Owing to the ability to measure proper motions in addition to LSR velocity, very long baseline interferometry (VLBI) maser 3D motions are used as a kinematic tracer for several astronomical phenomena, including tracing protostellar jets \citep[e.g.][]{Moscadelli19}, and expanding shells \citep{Moscadelli08}. These 3D motions can further be used to estimate the inclination of bipolar jets \citep{Sahai02}, the dynamical timescales of protostellar ejections \citep{Boekholt17}, and jet rotation \citep{Moscadelli08}. They are also used to estimate total 3D velocities of star forming regions (SFR) subsequently used in investigations of Galactic dynamics in the Milky Way, allowing substantially more accurate estimates of the Galactic constants i.e. the solar-Galactocentric distance $R_0$, Galactic circular rotation velocity $\Theta_0$, and the local standard of rest (LSR) angular rotation velocity $\Omega_0$. Despite the proficiency and current unique capability of VLBI maser observations for these purposes, a caveat remains in the assumption that maser 3D motions truly trace the motions of the phenomena to which they associate. However, there can be some discrepancy between the motions of individual maser cloudlets and the shocks or gas flows in which they reside. Another feasible complication to this assumption is the case where there exists a motion of the excitation conditions that give rise to maser emission rather than the gas itself, as was recently seen in the case of \qty{6.7}{GHz} methanol masers \citep{Moscadelli17,Burns20b}. 

There are considered to be two dominant pumping mechanisms that give rise to maser emission; those that are predominantly pumped by radiation, and those that are predominantly pumped by collisional energy. Radiatively pumped masers such as 6.7 GHz methanol masers are typically found near a heat source \citep{Zheng1997,Ouyang2019}, and are reasonably good tracers of motion of their hosting gas \citep{Green2011}. Cases of sudden changes to the radiative environment leading to drastic redistribution of such masers, as aforementioned, are exceptions. On the other hand, collisionally pumped masers such as water masers are usually found in shock regions \citep{Hollenbach13}. Such turbulent environments may give rise to situations where relative motions between gas flows, maser cloudlets in those gas flows, and the shock regions that provides the collisions required to pump the masers \citep{Strelnitsky1980,Strelnitski2002,Sobolev2003}. This scenario has the potential to lead to inaccurate estimates of jet motions, inclinations and dynamic timescales and the other maser-traced phenomena mentioned above.

Episodic accretion will trigger episodic ejections \citep{Corcoran1998}, and these ejections are visible via the formation of masers bow shocks which show the history of accretion events occuring during the formation of massive stars \citep{CarattioGaratti2015}. Tracing back the ejection history is an indirect way of investigating episodic accretion. This is done by measuring the velocity of these maser bow shocks allowing the deduction of its dynamic timescale which is basically the age of the ejection. 

Maser velocities have often been assumed to give a good approximation towards the jet velocities \citep{Goddi2005}. This study aims to evaluate the validity of the above assumption, which has important consequences on the use of maser VLBI determined dynamic timescales of ejections when inferring the accretion history of a high mass protostar.

Massive star-forming region, AFGL 5142 has an abundance of outflows on multiple scales \citep{Burns17}. It embeds two dominant 9 millimeter cores, MM1 and MM2, which exhibit hot core chemistry \citep{Zhang07}. MM1 has a mass of $\qty{6.5}{\mathrm{M}_\odot}$ \citep{Liu16} embeds a massive star traced by \qty{6.7}{GHz} methanol masers which detects an in-falling disk \citep{Goddi11}. The annual parallax of AFGL 5142 was measured to be $\pi = \qty{0.467\pm0.010}{mas}$, corresponding to a trigonometric distance of $D = \qty{2.14^{+0.051}_{-0.049}}{kpc}$ \citep{Burns17}. AFGL 5142-MM1 has a double bow shock structure which is rare in massive SFR and enables the evaluation of both the shock propagation and maser velocities. In addition, the source has experienced multiple previous episodes of protostellar ejections, which imply that the star has formed by episodic accretion \citep{Burns17}. The designated masers tracing the evolution of the previously detected North-West (NW) and South-East (SE) bowshocks from \citet{Burns17} were identified. In addition, some masers were detected further north from the bright NW bowshocks. Observations by \cite{Burns17} were conducted between 21 Apr 2014 and 29 May 2015, in 7 epochs.

\section{Observations}

Observations were carried out with the KVN and VERA array (KaVA) on 21st of March 2016 as part of the Star Formation Large Programme \citep{Kim17}, a united research program pioneered by the Korea Astronomy \& Space Science Institute (KASI) and the National Astronomical Observatories of Japan (NAOJ). The observations comprised alternating scans between the target maser, AFGL 5142-MM1 and a delay calibrator, DA193 at roughly hourly intervals. A total of 6 hours 20 minutes of KaVA observations were conducted, which included 1 hour 54 minutes of integration on AFGL 5142-MM1. The beams recorded signals are of left-hand circular polarization at all 7 KaVA observatories. Each station recorded data with a total bandwidth of \qty{256}{MHz} which was sampled into 16 baseband channels using the Octave system \citep{6051271}. We adopted \qty{22.235080}{GHz} as the rest frequency of the water maser. A radio frequency range of one baseband channel containing the maser emission was set carefully to cover all the maser spectral components based on calculations of Doppler shift at each observation date. 

Data were collected and correlated at Korea Japan Correlation Centre (KJCC) \citep{2015JKAS...48..125L}. Data from the Korean antennas did not produce useful correlated visibilities and were necessarily flagged. The phase tracking center for AFGL 5142-MM1 was set at $(\sigma,\delta)_{\text{J}2000.0} = (5^{\mathrm{h}}30^{\mathrm{m}}48^{\mathrm{s}}.000, +33^{\circ}47'54''.000)$. The \qty{16}{MHz} baseband channel which included the water maser emission was correlated with a channel separation of \qty{7.8125}{kHz} which corresponds to a velocity spacing of \qty{0.11}{km.s^{-1}}. 

\section{Data reduction}

The KaVA data was reduced via Astronomical Image Processing System (AIPS), established by National Radio Astronomy Observatory. Gain curve and system temperature information recorded at each station was used to apply a-priori gain calibration. Delay calibration was performed on a bright source, DA193, the solutions of which were then applied to the maser data. We solved residual phase and rate differences on each baseline using the AIPS task FRING. solving on one of the brightest channel in the maser emission (avoiding the brightest channel which exhibited complex structure), thus self-calibrating the spectral line data.

Maser emission was imaged using the AIPS task IMAGR, employing a pixel size of \qty{0.1 \times 0.1}{mas} and a map size of \qty{1024 \times 1024}{pixel}, which produced a map covering \qty{0.1024 \times 0.1024}{arcsec} centered on the area around the location of reference maser \citep{Burns17}. We also searched a $\qty{600 \times 600}{mas}$ region around the reference maser to ensure all maser emission was imaged. We produced an image cube from the velocity ranges of $\qtyrange{3.86}{-6.99}{km.s^{-1}}$. The image noise (rms) measured from an emission free region in a single channel map was $\qty{160}{mJy}$.



Water maser emission was elucidated from the image cube via the automated SAD routine in AIPS, which filters out candidate emission at a signal-to-noise ratio limit of 7. These `spots' represent the peak emission of an individual maser in a single channel, and they are grouped into `features' which infers to originate from the same physical maser cloud.

\section{Results}

\begin{figure}
    \centering
    \includegraphics[width=0.45\textwidth]{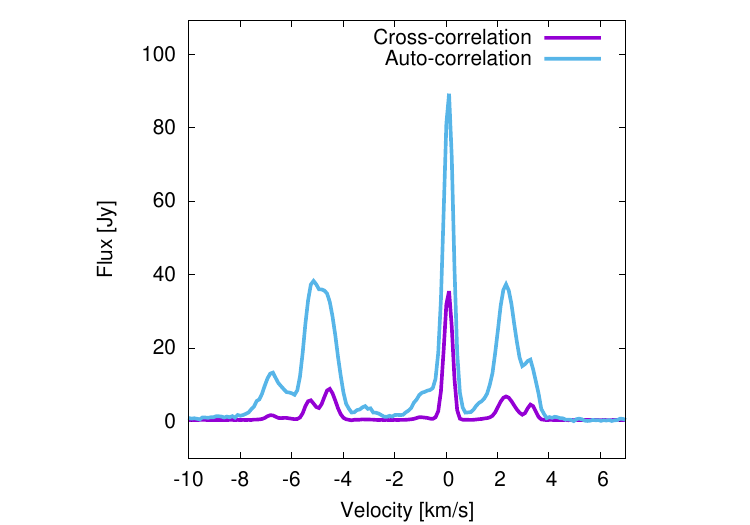}
    \caption{Cross-correlated and auto-correlated spectra of maser emission in AFGL 5142-MM1 observed using KaVA on the 21st of March 2016. The systematic velocity of the SFR is $\qty{-1.1}{km.s^{-1}}$.} 
    \label{fig:corr}
\end{figure}

\subsection{Spectral profile}

\begin{figure}
    \centering
    \includegraphics[width=.48\textwidth]{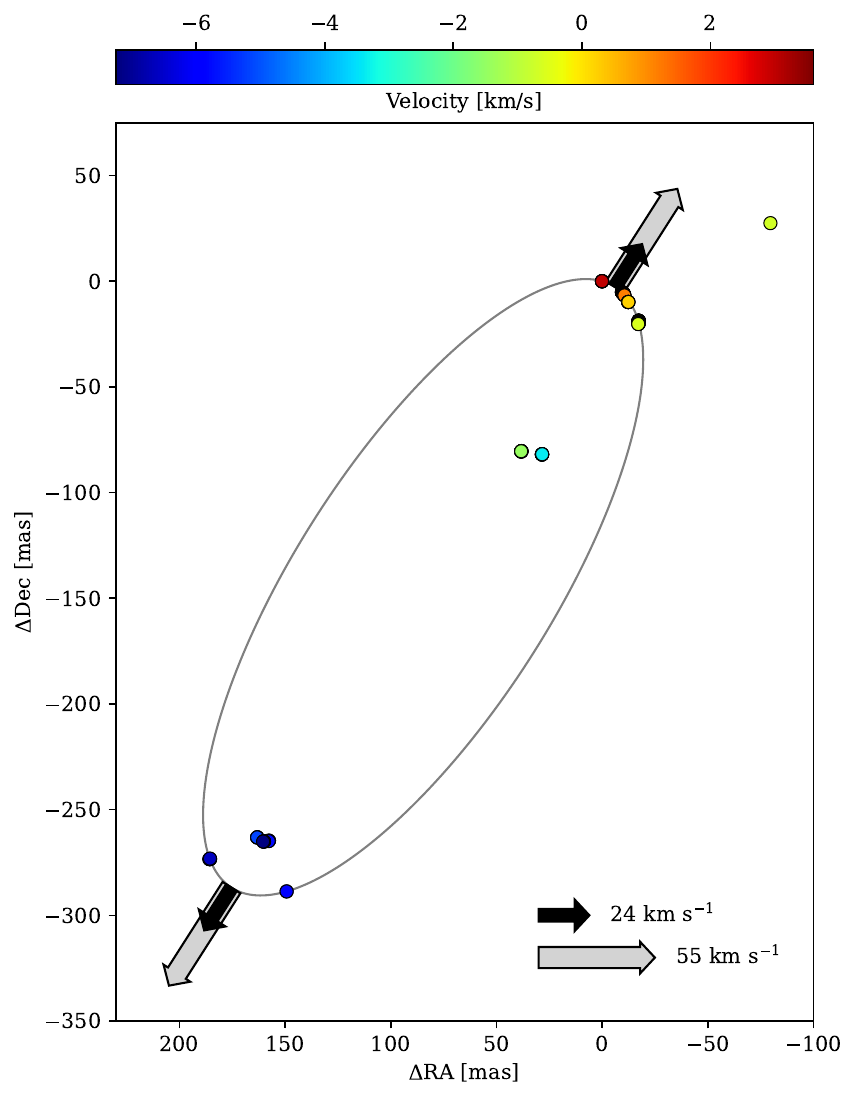}
    \caption{Spotmap of the maser emission in AFGL 5142-MM1 observed using KaVA on the 21st of March 2016. The grey ellipse shows the fit to the masers tracing the bowshock (see Section \ref{sec4.3}). The black vectors show the propagation speed of the bowshocks (this work, see Section \ref{sec5.2}) while grey vectors indicate the average proper motions of the maser spots associated with the same bowshocks, reported in \citet{Burns17}.}
    \label{fig:spotmap}
\end{figure}

Figure \ref{fig:corr} shows the detected flux density differences between cross-correlated and auto-correlated spectra of the water masers observed with KaVA on the 21st of March 2016, and Figure \ref{fig:spotmap} illustrates the spotmap of the water masers in AFGL 5142-MM1. The spectral profiles illustrate a high degree on consistency between both correlations, with more than 50\% of the maser emissions resolved out. This furthermore implies an abundance of information that could be explored in the extended scale ($\qty{\gg{1}}{mas}$) maser emissions in AFGL 5142-MM1.  

The KaVA observation on the 21st of March 2016 (See Figure \ref{fig:corr} and \ref{fig:spotmap}) shows that the detected water masers have velocities ranging from $\qtyrange{3.86}{-6.99}{km.s^{-1}}$. The l.o.s. velocity of the region was established as $\qty{-1.1}{km.s^{-1}}$ \citep{Zhang07} with the aforementioned NW and SE masers redshifted and blueshifted respectively \citep{Burns17}. This is consistent in Figure \ref{fig:corr} where the NW maser bowshocks are furthermore redshifted by approximately $\qty{4}{km.s^{-1}}$ and the SE masers by about $\qty{4}{km.s^{-1}}$ from the SFR systematic velocity. These clusters of maser bowshocks indicate symmetry in both the spatial and spectral domains.

 It should be noted that similarities in the spectra of this work and \cite{Burns17} indicate that masers associated with millimeter cores other than MM1 were detected by KaVA, however since they do not contribute to the primary goal of this work we did not process their data for imaging.

\subsection{Spatial distribution}

\begin{table*}
    \centering
    \begin{tabular}{cccccc}
        \hline
        & Center $x$ & Center $y$ & Rotation angle & Semi major axis & Semi minor axis \\
        & [\unit{mas}] & [\unit{mas}] & [\unit{rad}] & [\unit{mas}] & [\unit{mas}] \\
        \hline
        VERA fit & $61.6\pm0.2$ & $-101.0\pm0.2$ & $2.1789\pm0.0003$ & $166.7\pm0.2$ & $63\pm2$ \\
        Predicted & $71.2\pm0.5$ & $-98.0\pm0.5$ & $2.2545\pm0.0006$ & $171.1\pm0.4$ & $62\pm2$ \\
        KaVA fit & $84.65\pm0.06$ & $-144.78\pm0.08$ & $2.13877\pm0.00007$ & $168.58\pm0.04$ & $60.7\pm0.1$ \\
        \hline
    \end{tabular}
    \caption{Fit parameters for the ellipses for each set of data. The fitting was performed using numerically stable direct least squares, with the errors was estimated by minimization of $\chi^2$ function (further verified using bootstrapping).}
    \label{tab:ellipse}
\end{table*}

\begin{figure*}
    \centering
    \includegraphics[width=.7\textwidth]{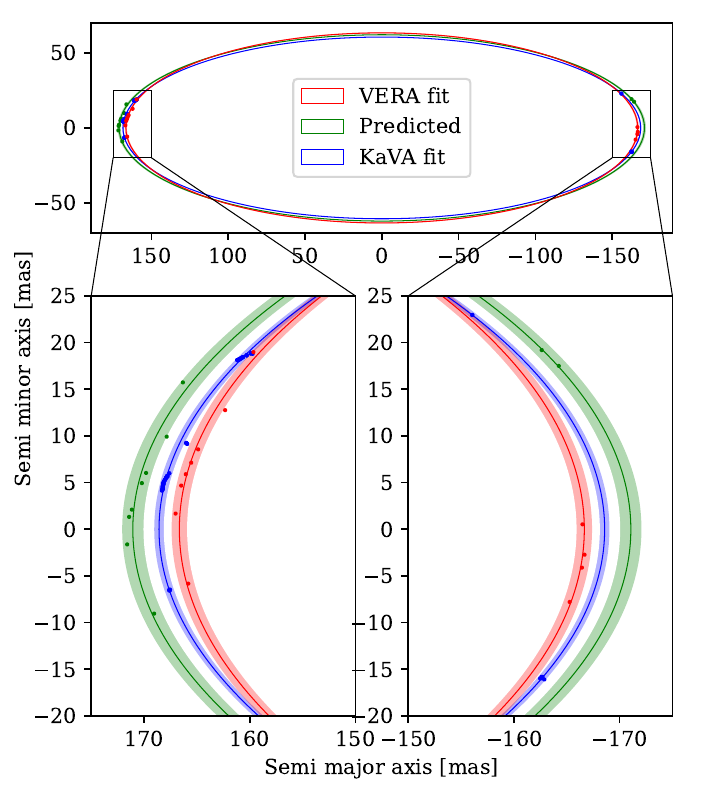}
    \caption{Comparison between the maser positions in VERA (May 2015), the predicted positions, and in KaVA (May 2016), along with their fitted ellipses. The positions in this plot are relative i.e. for each set of data, the maser positions are translated with respect to their center, then rotated about the center based on their rotation angle (original values shown in Table \ref{tab:ellipse}). The cutout in the lower left shows the NW region, and the cutout in the lower right shows the SE region. Note that any masers not associating with the bowshocks, and thus not included in this analysis, were omitted from the figure.}
    \label{fig:ellipse}
\end{figure*}

\begin{table}
    \centering
    \begin{tabular}{ccc}
        \hline
        & Separation \\
        & [\unit{mas}] \\
        \hline
        VERA fit & $333.3\pm0.4$ \\
        Predicted & $342.1\pm0.7$ \\
        KaVA fit & $337.17\pm0.07$ \\
        \hline
    \end{tabular}
    \caption{Separation between the NW and SE bowshocks and their errors for each set of data.}
    \label{tab:fitvalsanderr}
\end{table}

\begin{figure}
    \centering
    \includegraphics[width=0.45\textwidth]{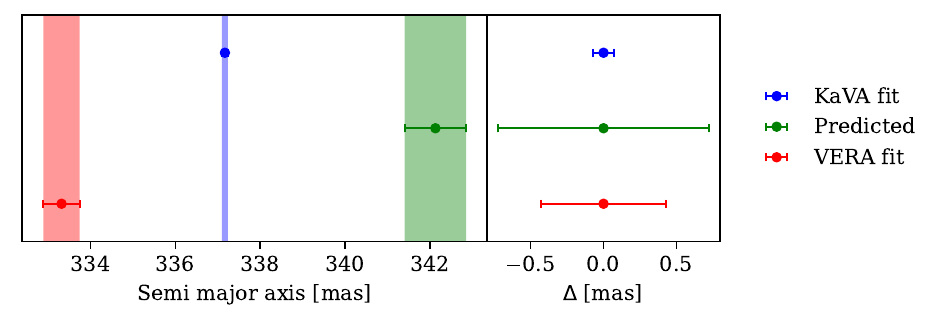}
    \caption{Visualization of the separation and their errors for each set of data. The left plot shows their values and error ranges, and right plot shows a comparison between each set's error values.}
    \label{fig:fitsvalsanderr}
\end{figure}

The masers associated with AFGL 5142-MM1 have an extent of about $\qty{300 \times 350}{mas}$, which is somewhat more extended than the distribution reported in \citet{Burns17}. 
 There were 91 maser spots detected, extended in a linear distribution orientated NW-SE. Most of the masers detected by KaVA associate with the known bowshock structures in this region. The presence of persistent and newly formed maser spots is apparent.
 
 While at first glance the maser distributions in this work resemble those from \citet{Burns17}, a closer look reveals that while the shock fronts themselves have persisted, the individual maser features have moved somewhat. This is evident in the blow up in Figure \ref{fig:ellipse} in the lower left and right cutouts. While some of the maser features appear to have persisted across the time between observations (2015 to 2016) several others have disappeared and new maser features have appeared in other parts of the shock fronts. This is not surprising since the shock driven water masers are known to have shorter lifetimes of a year or so \citep{Felli07} compared to the radiatively driven Class \textrm{II} methanol masers, for example. Thus, the shock front structure remains a persistent feature emanating from MM1, whereas the individual maser emitting regions are transient visitors that form, disappear, and reform in various parts of the shock front.

\subsection{Estimating bowshock displacement}
\label{sec4.3}

Our main objective in this work is to investigate whether the motion of the masers is a suitable tracer for the motion of the shock front itself. This can be determined by investigating whether or not the maser features are moving through the shock or with it. In the latter case, the locations of successive generations of maser features can be determined by the proper motion of the previous generation since both the masers and shock region itself should be co-moving. On the other hand, if the masers are moving through a shock region (in either a positive or negative velocity relation) subsequent generations of maser features will appear displaced from their expected locations based on the trajectories of the prior generation of masers. On one extreme, considering a static shock region, individual masers will exhibit a proper motion, however long term VLBI monitoring will reveal that subsequent generations of masers are consistently produced at the same location before traversing the shock region and eventually disappearing. Given that maser emission traces shock regions, and given enough time between observations for a new generation of maser features to be produced, the two scenarios above can distinguished by a comparison between the translocation of the shock structures over the given time period, and their expected translocation based on the proper motions of masers they host.

To investigate any discrepancy between the predicted and actual locations of the shock regions we produced a synthetic spotmap in which the maser spots observed in May 2015 were extrapolated to their expected positions to March 2016, based on their measured proper motions from \citet{Burns17}. This represents the expected location of the shock front in March 2016 assuming that the maser and the shock are co-moving. We then compared this to the data taken in this work which gives the actual location of the shock front in March 2016. The separation distance of the shock fronts then reveals how far they have propagated in the sky plane in the given time.

In order to determine how far apart the NW and SE maser shock fronts have separated between the VERA (May 2015) and KaVA (March 2016) observations, in addition to their expected separation based on proper motion trajectories, we fit ellipses to the maser spot distribution using an modified version of the procedure used in \cite{Burns17}. Since the KaVA observations did not use phase referencing we fit ellipses to each data set to derive its center, and then shifted each data set to a common origin. After doing so the separation of the NW and SE bowshocks were determined by the semi-major axis of the ellipses for each corresponding spotmap, and the motion in each direction is half of the semi-major axis expansion as a function of time. Comparing the separation of the maser bowshocks as a function of time avoids the complication of considering the source systemic motion in the sky plane, thereby reducing the number of variables.

The results are as follows. The separation between the NW and SE shock regions was \textbf{$\qty{333.3\pm0.6}{mas}$} in May 2015. Based on maser trajectories, the separation in March 2016 was expected to be \textbf{$\qty{342\pm1.4}{mas}$}. The separation of the bowshocks in this work was determined to be \textbf{$\qty{337.2\pm0.2}{mas}$}. In Figure \ref{fig:ellipse} we show an expansion of the maser distribution in the NW shock front showing the maser distributions and the determined shock region locations represented by the fitted ellipses (shown in gray with a width corresponding to the typical astrometric error adopted for VERA and KaVA observational results \citep{Niinuma15}.

 It can be seen that the shock front detected via KaVA in these observations, delineated by the locations of a new generation of masers, does not move as far as is expected (predicted locations) in the 10 months between the two VLBI data sets.

\section{Discussion}

\subsection{Maser proper motion vs shock propagation}

Multi-epoch VLBI observations can trace the proper motions of maser emission in the sky plane. In conjunction with this, consideration of the LSR velocity gives a 3D description of the spatial motion of the masers. This technique has been utilized in tracing the dynamics of star formation from the parameters obtained via analysis of ejections (see, for example, \citealt{Goddi2005}).

 It is assumed that the measured motions of masers accurately trace the motion of the maser emitting gas, which in the case of shock-tracing water masers would presume that the 3D motion of the masers would be equal to the 3D motion of the shock region. Theoretically, we could obtain physical quantities such as the jet inclination from this information by the ratio of the line-of-sight and sky-plane velocities. The jet inclination is crucial in understanding the production and dynamical ages of protostellar outflows in addition to determining the rotation axis inclination of accretion disks. A few hypotheses for jets in massive young stellar objects include an extension of low-mass protostars as classical-disk wind, and/or external collimation of a wide-angle wind \citep{CarrascoGonzlez2021}. 

Our results have shown that discrepancies may exist between the motions of the shock and the motions of the masers. This brings into question this assumption i.e. masers may not always precisely trace the motions of the shock. This may be due to a relative motion between the propagating shock and the maser cloudlets propagating with or within the shock region, or supersonic turbulence that occurs in the gas after the shock has left the cloudlets, which gives rise to collisionally pumped maser emissions \citep{Strelnitski2002} Ultimately, this observed effect can be attributed to the changing pumping conditions within the shock, and this depends on the physical environment and thus could vary for different sources

It should be stressed that we do not negate the utility of masers as an estimate of the geometric and kinematic attributes of shocks but rather highlight a limitation in this approach. Indeed, there are confirmed cases where the shock motion and the motions of associated maser cloudlets are consistent \citep{Hirota21a, Goddi07, Burns16a}. However, the determination of any corresponding parameters such as ejection rates, based on the proper motions of the masers should be done with caution. Increasing the number of motion comparison studies with observational time baselines of decades would confirm our findings to a higher degree of confidence, in addition to exploring the question of why some masers do or do not trace well to the motion of the shock regions in which they form. Such VLBI monitoring investigations would allow ample time for several generations of water masers to form, traverse, disperse, and be replaced by subsequent generations of masers. Our hope is to build upon the understanding of this topic by extending and repeating the study presented in this publication.

\subsection{Deceleration of shocks}
\label{sec5.2}

Another perspective to look into based on our results is the fact that the shock itself may slow down. It has been known that the medium around MYSOs is inhomogeneous, with one of the strongest evidence being the presence of spiral arms in their accretion disk \citep{Burns2023}. Molecular line observations such as HNCO also agree with the presence of denser regions in molecular outflows or shocks \citep{Xie2023}. The shock would naturally slow down when it meets a denser part of the circumstellar medium.

Multiple ejection events can also produce the same result. Bow shocks from previous ejections can excavate an outflow cavern to produce a less dense region in the circumstellar medium. A secondary shock can then propagate faster in this region, approaching the slower initial/primary bow shocks. The secondary shock would then slow down as it interacts with the primary bow shocks \citep{Meyer2020}. However, considering that the physical difference between the old and new masing fronts is only \qty{4}{au}, it may be more reasonable to interpret our results as shock front propagation rather than multiple ejections.

The presence of multiple jets which are linked to jet precession can also produce wide-angle shocks with inconsistent velocities, but this should affect the maser spatial distribution rather than its velocities, and its effects in AFGL 5142 has only been seen at a much larger scale \citep{Liu16}.

Non-linear velocities are also seen in different astrophysical objects, namely the BX Cam envelope, a circumstellar envelope around a Mira variable star \citep{Xu2022}. Water maser velocities in the source was seen to not be constant based on the asymmetric spatial and velocity distributions in the maser features. The authors managed to quantify maser acceleration too, albeit the values are within uncertainties.

Nonetheless, the shock velocities predicted based on the 3D velocities using the VERA observation is \qty{55\pm5}{km.s^{-1}}, whereas the KaVA data corresponds to a shock velocity of \qty{24\pm3}{km.s^{-1}}.

\section{Conclusion}

The short-term positional monitoring, time scale lifespan of the water maser, will give a lower limit on the shock velocity. Then the long-term monitoring (several life spans of water masers) will reveal the true motion of the shock. If both reveal the same velocity, hence both the shock and the jet are moving together. 

While water masers are useful tools for tracing the spatial kinematics of jets. In protostars, caution should be exercised when making assumptions on how accurately/closely the 3D kinematics of masers observed via VLBI trace the true physical motion of the shock regions in the jet.

We conclude that the 3D velocity of masers may not accurately give all information about the jet. Short-term and long-term VLBI monitoring will investigate different parts of the outflow, either the maser themselves or the shock. This work has compared both short-term (VERA) and long-term (KaVA) observations and our theory holds true.

\section*{Acknowledgements}

Z.R. acknowledges the support provided by UM's Faculty of Science grant (GPF081A-2020), which has enabled the completion of this research and the publication of its findings. T.H. is supported by the MEXT/JSPS KAKENHI Grant Numbers 17K05398 and 20H05845. T.L. acknowledges supports from the National Key Research and Development Program of China (No. 2022YFA1603101); the National Natural Science Foundation of China (NSFC), through grants No. 12073061 and No. 12122307; the international partnership program of the Chinese Academy of Sciences, through grant No. 114231KYSB20200009; and the Shanghai Pujiang Program 20PJ1415500. C.W.L. is supported by the Basic Science Research Program through the National Research Foundation of Korea (NRF) funded by the Ministry of Education, Science and Technology (NRF-2019R1A2C1010851), and by the Korea Astronomy and Space Science Institute grant funded by the Korea government (MSIT) (Project No. 2023-1-84000). A.M.S. was supported by Ministry of Science and Higher Education of the Russian Federation (state contract FEUZ-2023-0019).





\bibliographystyle{mnras}
\bibliography{final} 


\bsp	
\label{lastpage}
\end{document}